\documentclass[%
 reprint,
 amsmath,amssymb,
 aps,
]{revtex4-1}

\usepackage{graphicx}
\usepackage{dcolumn}
\usepackage{bm}
\usepackage{graphicx}
\usepackage{dcolumn}
\usepackage{bm}
\usepackage{bigints}
\usepackage{color}
\usepackage{amsmath,amssymb,graphicx}
\usepackage{float}
\usepackage{braket}
\usepackage{amsmath}
\usepackage{textcomp}

\begin{document}

\preprint{APS/123-QED}

\title{Laser Cooling Characterization of Yb-Doped ZBLAN Fiber as a Platform for Radiation Balanced Lasers}

\author{Mostafa Peysokhan$^{1,2}$}
\author{Esmaeil Mobini$^{1,2}$}
\author{Arman Allahverdi$^{2,3}$}
\author{Behnam Abaie$^{1,2}$}
\author{Arash Mafi$^{1,2,*}$}
\affiliation{$^1$Department of Physics \& Astronomy\\
             $^2$Center for High Technology Materials\\
             $^3$Department of Electrical and Computer Engineering, University of New Mexico, Albuquerque, New Mexico 87131, USA\\
             $^*$Corresponding author: mafi@unm.edu}%

\date{\today}

\begin{abstract}
Recent advances in power scaling of fiber lasers are hindered by the thermal issues, which deteriorate the beam quality. Anti-Stokes fluorescence cooling has been suggested as a viable method to balance the heat generated by the quantum defect and background absorption. Such radiation-balanced configurations rely on the availability of cooling-grade rare-earth-doped gain materials. Herein, we perform a series of tests on a ytterbium-doped ZBLAN optical fiber to extract its laser cooling-related parameters and show that it is a viable laser cooling medium for radiation-balancing. In particular, a detailed Laser Induced Modulation Spectrum (LITMoS) test is performed to highlight the transition of this fiber to the cooling regime as a function of the pump laser wavelength. Numerical simulations support the feasibility of a radiation-balanced laser, but highlight that practical radiation-balanced designs are more demanding on the fiber material properties, especially on the background absorption, than are solid-state laser cooling experiments.
\end{abstract}

\pacs{Valid PACS appear here}

\maketitle
\section{Introduction}
Fiber lasers are attractive sources of high-power coherent radiation for industrial and directed energy applications. 
They have many excellent properties; e.g., they enjoy high power efficiencies, broad gain linewidths, and diffraction limited beam qualities. 
Moreover, the availability of fully fiberized cavities without the need for precise alignment makes them quite flexible for 
implementation outside the controlled environment of research laboratories~\cite{richardson2010high,zervas2014high}. 
Efficient heat mitigation plays an important role in the quest to achieve ever-increasing output powers from fiber lasers and amplifiers. 
Current approaches to power scaling are limited by the thermally-induced mode instability, which degrades the output beam 
quality~\cite{brown2001thermal, zenteno1993high, ward2012origin, dawson2008analysis, jauregui2012physical}. Anti-Stokes fluorescence (ASF) cooling has 
been suggested as a viable method to address such thermal issues~\cite{epstein1995observation, seletskiy2010laser}. 
In practical designs, ASF cooling can reduce the heat-load or even balance the heat generated by the quantum defect and background 
absorption, which is dubbed as radiation-balancing~\cite{bowman1999lasers,bowman2010minimizing,bowman2016low,yang2019radiation}.

The development of a viable radiation-balanced Yb-doped fiber laser depends on the synthesis of a high-purity cooling-grade
glass host. At this point, the most viable material to demonstrate the concept of a radiation-balanced laser (RBL) in an optical fiber platform
is the Yd-doped ZBLAN glass (${\rm ZrF}_4$-${\rm BaF}_2$-${\rm LaF}_3$-${\rm AlF}_3$-${\rm NaF}$), because it has been successfully cooled via ASF~\cite{epstein1995observation}. 
It also has interesting properties for certain specialized high-power applications, especially because ZBLAN can be doped with a higher Yb concentration 
than silica glass~\cite{PhysRevApplied.11.014066}. The relatively small quantum defect of Yb dopants, high doping concentration, 
wide pump absorption band, availability of low-cost and high-brightness pump diodes in the 0.9-1.0\,\textmu m absorption band of Yb, and the possibility of making
a double-cladding fiber configuration make Yb-doped ZBLAN fibers an attractive choice for fiber RBL operation~\cite{mungan1997laser, gosnell1999laser,knall2019demonstration}.

In the quest to develop a viable radiation-balanced fiber laser or amplifier, it is important to fully characterize the optical (gain and absorption) and ASF cooling properties of 
Yb-doped ZBLAN glass fibers. In particular, the wavelength dependence of the ASF cooling efficiency is needed for choosing the optimal pump and laser
wavelengths. Moreover, it has been shown that the heat generation due to the parasitic absorption of the host glass plays an important role in setting the
thermal behavior of a high-power fiber laser system and can even dominate the heat generation due to the quantum defect~\cite{mobini2018thermal}. Therefore, it is critical to accurately determine
the parasitic absorption of Yb-doped ZBLAN glass fibers. In this paper, we combine two techniques: the ``Laser Induced Modulation Spectrum'' (LITMoS) test
developed in Sheik-Bahae's research group~\cite{melgaard2014identification}, 
and our recently developed ``Measuring the Absorption Coefficient via Side-Light Analysis'' (MACSLA) method~\cite{peysokhan2019method} to characterize 
the cooling efficiency of a cooling-grade Yb-doped ZBLAN glass fiber as a function of the wavelength and determine its resonant and parasitic absorption 
properties. We will then use the extracted parameters to explore the design and optimization of a Yb-doped ZBLAN glass fiber laser system.   

To quantify the ASF cooling, it is common to use the cooling efficiency, $\eta_c$, of which determination is a primary focus of this paper. 
In a conventional laser cooling setup, where the material is pumped by a laser at frequency $\nu_p$ and cooled via ASF,  
$\eta_c$ is the net power density (per unit volume) extracted from the material ($P_{\rm net}$) per unit power density absorbed or scattered 
($P_{\rm abs}$): $\eta_c=P_{\rm net}/P_{\rm abs}$. It can be shown that the cooling efficiency can be written as
\begin{align}
\label{eq:etac}
\eta_c(\lambda_p)=\dfrac{\lambda_p}{\lambda_f}\,\eta_q\,\eta_{\rm abs}(\lambda_p)-1,
\end{align}
where the absorption efficiency is given by
\begin{align}
\label{eq:etaabs}
\eta_{\rm abs}(\lambda_p)=\dfrac{\alpha_r(\lambda_p)}{\alpha_r(\lambda_p)+\alpha_b}.
\end{align}
$\alpha_r(\lambda_p)$ is the resonant pump absorption coefficient and $\alpha_b$ is the parasitic background absorption coefficient.
$\lambda_f$ is the mean fluorescence wavelength,  $\eta_q$ is the external quantum efficiency, and $\eta_{\rm abs}(\lambda_p)$ is the absorption 
efficiency at the pump wavelength $\lambda_p=c/\nu_p$ ($c$ is the speed of light in vacuum). We present a derivation of Eq.~\ref{eq:etac} in the Appendix,
but for here, it is sufficient to know that $0\le \eta_q\le 1$ and $0\le \eta_{\rm abs}\le 1$.

As already mentioned, determining the wavelength-dependence of the cooling efficiency, $\eta_c$, is a main focus of this paper. For ASF cooling $P_{\rm net}$
must be positive (net heat extraction), which is equivalent to a positive value for the cooling efficiency, $\eta_c$. From Eq.~\ref{eq:etac}, one can immediately determine that 
$\lambda_f<\lambda_p$ is a necessary condition for ASF cooling, given that $0\le \eta_q, \eta_{\rm abs}\le 1$. In practice, both $\eta_q$ and $\eta_{abs}$
must be very close to unity to observe ASF cooling. The reason is that $\lambda_p$ cannot be much longer than $\lambda_f$, otherwise the resonant pump absorption 
coefficient, $\alpha_r(\lambda_p)$, would become too small, hence lowering the value of the absorption efficiency $\eta_{\rm abs}(\lambda_p)$ (see Eq.~\ref{eq:etaabs}). 
Therefore, in an ASF cooling experiment, $\lambda_p/\lambda_f$ ($\eta_{\rm abs}(\lambda_p)$) is a monotonically increasing (decreasing) function of the pump wavelength; it 
is the balance between $\lambda_p/\lambda_f$ and $\eta_{\rm abs}(\lambda_p)$ if $\eta_q\approx 1$, which dictates an ASF cooling range in $\lambda_p$.
In the following, we will use the LITMoS test and measure the wavelength dependence of $\eta_{\rm abs}$, which allows us to determine $\eta_q$ and $\eta_{\rm abs}$ 
for the cooling-grade fiber. Once $\eta_{\rm abs}$ is determined, we will then apply the MACSLA method to find the values of the parasitic absorption $\alpha_b$, 
as well as $\alpha_r(\lambda_p)$.  
\section{Wavelength dependence of the cooling efficiency}
\label{sec:CE}
\begin{figure}[h]
\begin{centering}
\includegraphics[width=3 in]{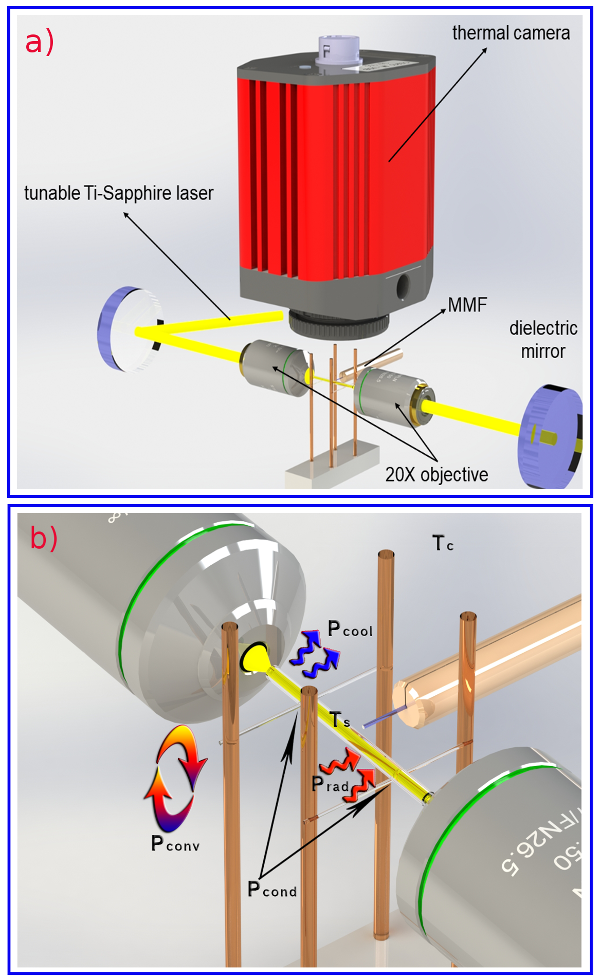}
\caption{a) Experimental setup for the LITMoS test of the Yb-doped ZBLAN fiber. b) Magnified image of the 
fiber holder and an illustration of the three sources of heat load: convective, conductive and radiative.}
\label{fig:litsetup}
\end{centering}
\end{figure}

The experimental setup to perform the LITMoS test and measure the wavelength dependence of the cooling efficiency, $\eta_{\rm abs}$, is depicted in Fig.~\ref{fig:litsetup}. 
All the measurements are done for a multimode ($1\%\,{\rm YbF}_3$) ZBLAN fiber with the core diameter of 300\,\textmu m and cladding diameter of 430\,\textmu m.
Both facets of the fiber are polished with the cooling grade polishing technique that is detailed in the Methods section. 
Per Fig.~\ref{fig:litsetup}a, the ZBLAN fiber is pumped from the left-side by a tunable Ti:Sapphire laser, which is coupled to the ZBLAN fiber via a 20x microscope objective. 
The pump is reflected back into the fiber by an objective-mirror combination in a double-pass configuration. The fiber temperature is measured by a thermal camera, which is 
placed on top of the ZBLAN fiber. To minimize the thermal interaction between the ZBLAN fiber and the environment, the ZBLAN fiber is supported by two thin glass fibers 
as shown in Fig.~\ref{fig:litsetup}b.  

The fluorescence spectral power density of the ZBLAN fiber, $S(\lambda)$, is measured from the side by a multimode optical fiber, which is coupled to a spectrometer.  
We note that the position of this multimode optical fiber does not change relative to the ZBALN fiber during the LITMoS test. Therefore, the absorbed power density, $P_{\rm abs}$ at 
each pump wavelength, $\lambda_p$, is proportional to the total collected fluorescence spectral power. In other words, $P_{\rm abs}(\lambda_p)\propto \int d\lambda\,S_p(\lambda)$, where 
the integral is performed over the entire fluorescence spectrum, and the subscript $p$ in $S_p(\lambda)$ signifies that the fluorescence spectral power density relates to pumping 
the ZBLAN fiber at $\lambda_p$. We emphasize that by changing $\lambda_p$, only the overall intensity of $S_p(\lambda)$ is re-scaled and its spectral form does not change. 
$P_{\rm net}$ is proportional to the change in the temperature, $\Delta T$, of the fiber, which is proportional to $\Delta({\rm pixel})$ of the image captured by 
the thermal camera at each $\lambda_p$. Therefore, for the pump wavelength, $\lambda_p$, the cooling efficiency is approximated by $\eta_c\propto -\Delta({\rm pixel})/\int d\lambda\,S_p(\lambda)$. 

\begin{figure}[htp]
\begin{centering}
\includegraphics[width=3.2 in]{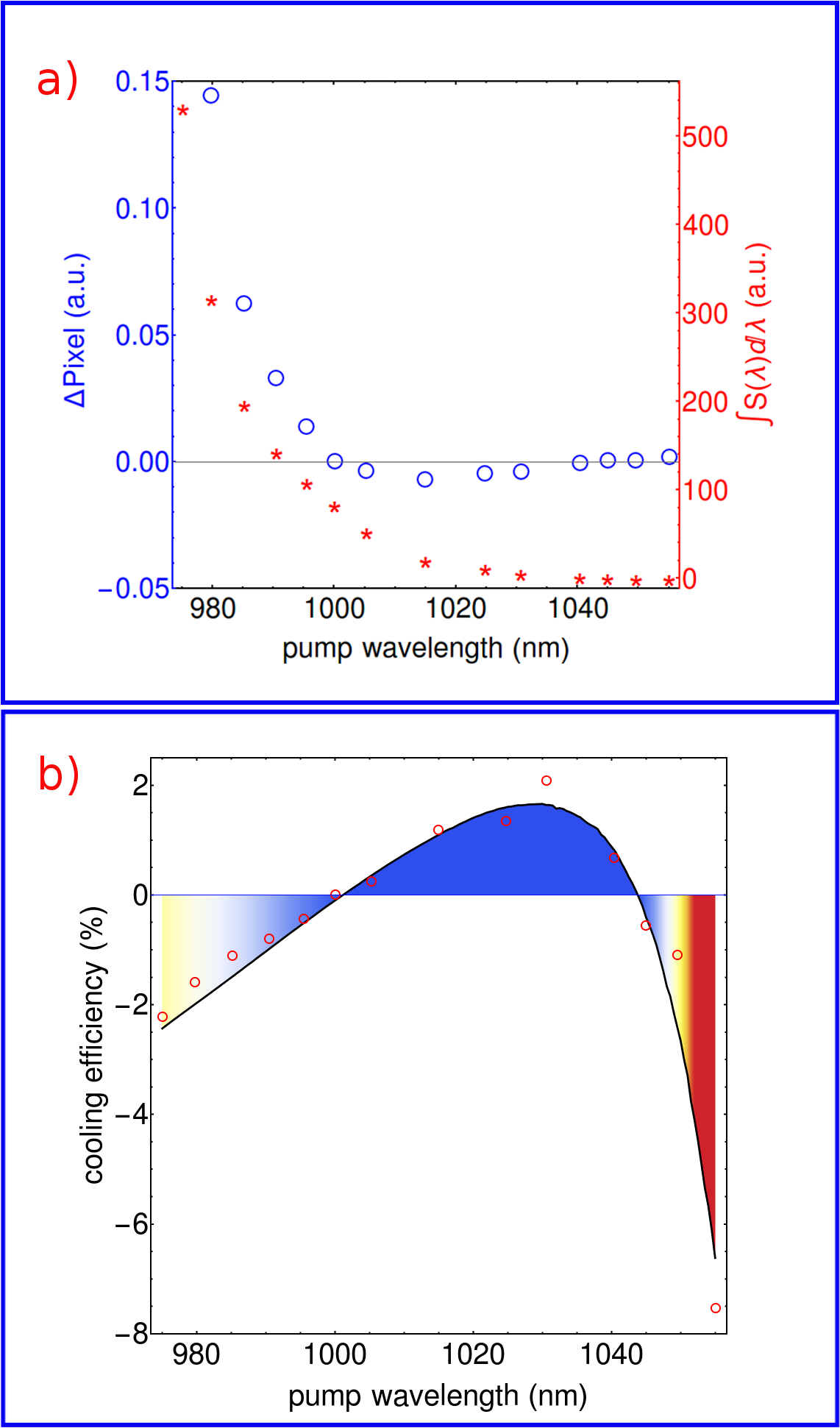}
\caption{a) Blue circles correspond to $\Delta({\rm pixel})$ (change in the pixel value of the thermal camera image) at each wavelength, 
and red asterisks represent the area under the $S(\lambda)$ curve. b) Red dots represent the measurement of the cooling efficiency ($\eta_c$) for the Yb:ZBLAN fiber
at different wavelengths. The solid curve shows a fitting of $\eta_c$ based on Eq.~\ref{eq:etac} to the measured values, where the positive region in $\eta_c$ indicates cooling.}
\label{fig:litdata}
\end{centering}
\end{figure}
In Fig.~\ref{fig:litdata}a, the blue circles correspond to $\Delta({\rm pixel})$ (change in the pixel value of the thermal camera image) 
and the red asterisks represent $\int d\lambda\,S_p(\lambda)$. The ratio is plotted 
in Fig.~\ref{fig:litdata}b and is fitted to Eq.~\ref{eq:etac}, where the $-\Delta({\rm pixel})/\int d\lambda\,S_p(\lambda)$ ratio is renormalized by a single overall 
scaling factor to conform to Eq.~\ref{eq:etac}. We note that $\alpha_r(\lambda_p)$ follows a strict spectral function of the 
form~\cite{PhysRevApplied.11.014066, newell2007temperature, aull1982vibronic}:
\begin{align}
\alpha_r(\lambda) \propto \lambda^5\, S(\lambda)\, \exp\left(\dfrac{hc}{\lambda k_B T}\right), 
\label{Eq:abs}
\end{align}
which is used in Eq.~\ref{eq:etac} and Eq.~\ref{eq:etaabs} to perform the fit. Here, $h$ is the Planck constant, $k_B$ is the Boltzmann constant, 
$c$ is the speed of light in vacuum, and $T$ can be the room temperature as long as the temperature variation due to ASF cooling is not large.
Equation~\ref{Eq:abs} allows us to replace $\alpha_r(\lambda)$ in Eq.~\ref{eq:etaabs} with $\alpha^p_r\times \widetilde{\alpha}_r(\lambda)$, 
where $\widetilde{\alpha}_r(\lambda)$ is the absorption coefficient normalized to its peak value, $\alpha^p_r=\alpha_r(\lambda_{\rm peak})$. 
Therefore, the fitting procedure in Fig.~\ref{fig:litdata}b becomes a two-parameter fit (besides the overall scaling), to determine the ratio $\alpha_b/\alpha^p_r$ 
and the external quantum efficiency, $\eta_q$. We find that $\alpha_b/\alpha^p_r = 2.363 \times 10^{-4}$ and $\eta_q=99.6\%$.
\section{Measuring the resonance absorption}
\label{sec:RA}
In the previous section, we managed to determine the external quantum efficiency, $\eta_q$, along with the ratio of the parasitic background absorption to the peak resonant absorption,
$\alpha_b/\alpha^p_r$. In order to find the actual values of $\alpha_b$ and $\alpha^p_r$ (not just the ratio), we can now use the 
MACSLA method~\cite{peysokhan2019method, peysokhan2019non,peysokhan2018non}. 
The MACSLA method is based on comparing the collected spontaneous emission power at two arbitrary points along the fiber for different pump wavelengths. 
For a multimode optical fiber, due to a larger Yb-doped core diameter and consequently a stronger signal from the side of the fiber, it is not necessary to use a lock-in amplifier 
that was detailed in Ref.~\cite{peysokhan2019method} for a single-mode implementation. Here, we measured the spontaneous emission power directly by 
a power meter from the side of the fiber. Because the Yb-doped ZBLAN fiber is multimode, we used a passive multimode optical fiber to fully scramble the pump modes 
before coupling the pump to the core of the active fiber to improve the pump absorption efficiency.

\begin{figure}[h]
\begin{centering}
\includegraphics[width=3 in]{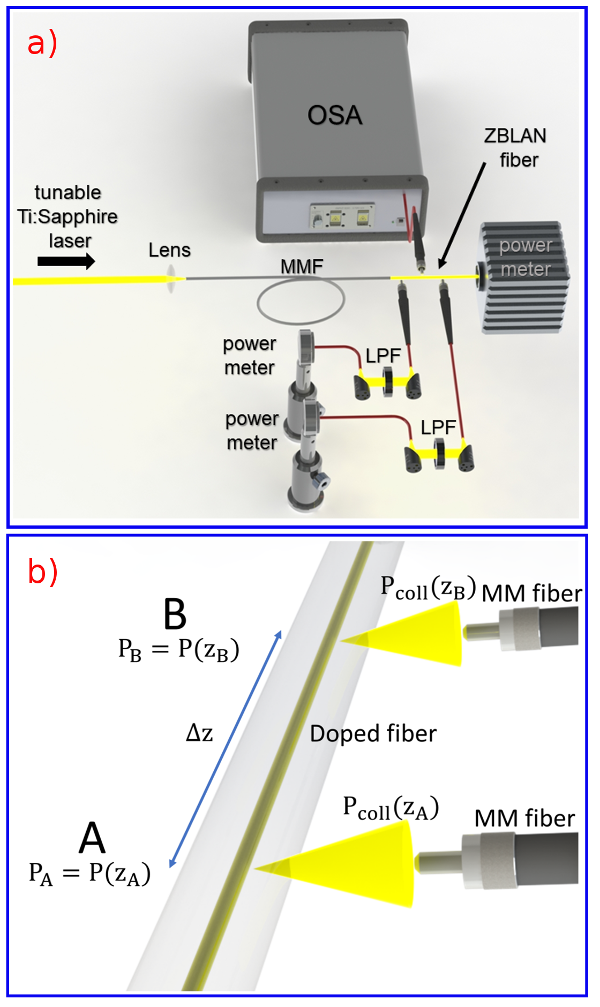}
\caption{a) Schematic of the experimental setup which is used for the MACSLA method. OSA stands for optical spectrum analyzer, LPF for long-pass filter, 
and MMF for multimode fiber. b) Schematic of the propagation of the pump power in the core of the optical fiber, and collection of the spontaneous emission 
from the side of the Yb-doped ZBLAN fiber by two multimode passive optical fibers.}
\label{fig:macsla}
\end{centering}
\end{figure}

The experimental setup is shown in Fig.~\ref{fig:macsla}(a), where a Tunable Ti:Sapphire laser beam is coupled to a multimode fiber (infinicore 300, Corning) with the 
length of 3\,m and the output of the multimode fiber is butt-coupled to the ZBLAN fiber. Two other multimode fibers (M124L02, Thorlabs) are employed to collect the 
spontaneous emission from the side of the doped fiber at two different locations, points {\bf A} and {\bf B} marked by positions $z_A$ and $z_B$, respectively, 
alongside the ZBLAN fiber. The collected side light is filtered with a 1.0\,\textmu m long-pass filter to remove the scattered pump and the filtered collected 
fluorescence is measured with a sensitive power meter (S120C, Thorlabs). Figure~\ref{fig:macsla}(b) shows a schematic of the MACSLA method. 

\begin{figure}[h]
\begin{centering}
\includegraphics[width=3.2 in]{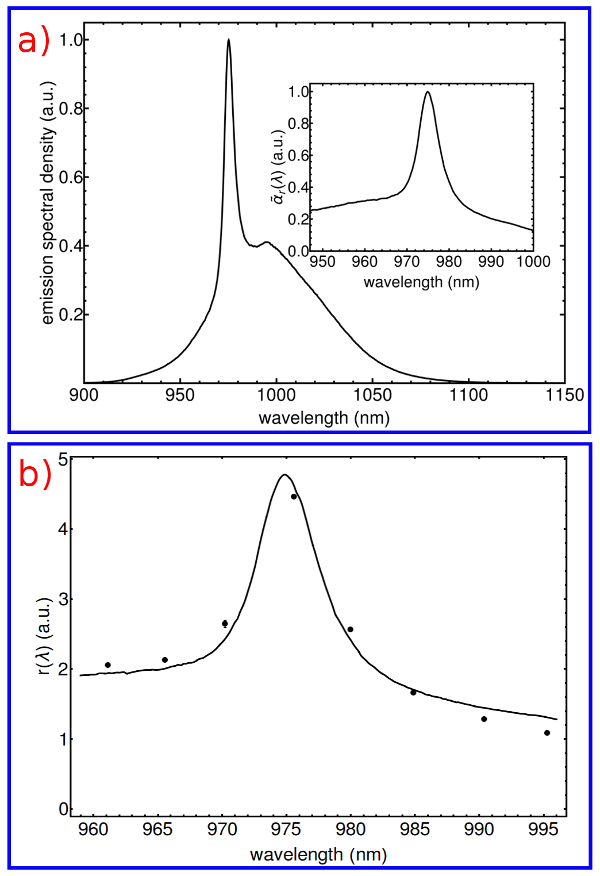}
\caption{a) Emission power spectral density $S(\lambda)$, which is measured by the optical spectrum analyzer, is plotted in arbitrary units.
The inset shows the resonant absorption coefficient, which is normalized to its peak value and is calculated by using the McCumber theory. 
b) The points indicate the values of $r(\lambda)$ measured at different wavelengths near the peak of the resonant absorption coefficient.}
\label{fig:mm}
\end{centering}
\end{figure}

The power spectral density $S(\lambda)$ of the Yb-ZBLAN fiber is shown in Fig.~\ref{fig:mm}(a). The inset shows the resonant absorption coefficient, 
which is normalized to its peak value and is calculated by using the McCumber theory~\cite{mccumber1964einstein}. The fitted line to the experimental 
measurements related to $r(\lambda)\,=\,\ln\left[P_{\rm coll}(z_B)/P_{\rm coll}(z_A)\right]$ is shown in Fig.~\ref{fig:mm}(b). The points 
indicate the values of $r(\lambda)$ measured at eight different wavelengths, and the fitting curve comes directly from the resonant absorption spectrum 
shown as the inset in Fig.~\ref{fig:mm}(a). The final outcome of the fitting process is the peak value of the resonant absorption coefficient, $\alpha^p_r\,=\,1.86\,{\rm cm}^{-1}$,
which can be combined with the result form the LITMoS test to give $\alpha_b=4.278 \times 10^{-2} {\rm m}^{-1}$.

The results presented in sections~\ref{sec:CE} and~\ref{sec:RA} amount to a detailed laser-cooling characterization of the Yb-doped ZBLAN optical fiber using the LITMoS and MACSLA methods.
In particular, we have captured the wavelength dependence of the cooling efficiency $\eta_c$, the value of the external quantum efficiency $\eta_q$, 
the peak value and wavelength dependence of the resonance absorption coefficient $\alpha_r(\lambda)$, as well as the value of the parasitic background absorption $\alpha_b$.
The accurate determination of these parameters is essential for properly designing a radiation-balanced fiber laser or amplifier, an example of which will be done in the next section   
\section{Simulation and the results}
\begin{figure}[h]
\begin{centering}
\includegraphics[width=\columnwidth]{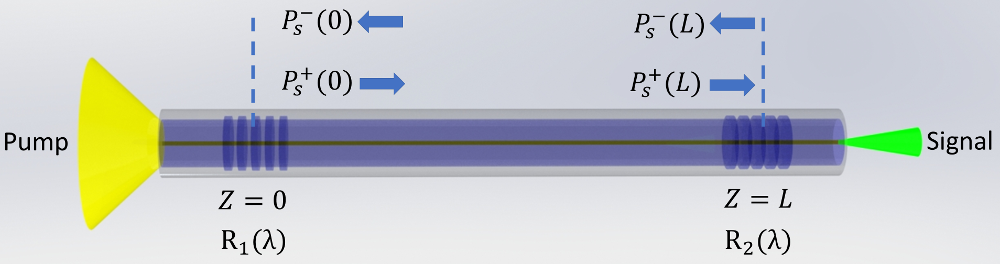}
\caption{Schematic of the laser system and propagation of the pump power and signal in the double-cladding fiber laser. 
Pump power is launched at $z=0$ and the output signal is calculated at $z=L$ at the power delivery port. $R_1(\lambda)$ and $R_2(\lambda)$ are the 
distributed Bragg reflectors at $z=0$ and $z=L$.}
\label{fig:laser}
\end{centering}
\end{figure}

In this section, we use the experimental results on the detailed characterization of the Yb-doped ZBLAN fiber from sections~\ref{sec:CE} and~\ref{sec:RA} 
to investigate the possibility of designing a viable radiation-balanced fiber laser. Our simulations are intended to highlight the possibility of designing 
such lasers with negligible heat production to address the laser heating and mode-instability problems. For our simulations, we use the main cooling-related 
parameters such as the cooling efficiency as a function of the wavelength, resonance absorption coefficient, background absorption coefficient, 
and the external quantum efficiency, all of which were found experimentally in the previous sections. The platform for our design is a double-cladding fiber 
laser geometry with two distributed Bragg reflectors on each side of the fiber as shown in Fig.~\ref{fig:laser}, which is the common platform for high-power operation.

We consider an active fiber of length $L$ with the dopant density $N$, which is assumed to be constant along the fiber (independent of the position $z$). 
The pump power ${P_p}(z)$ at wavelength $\lambda_p$ is coupled into the inner cladding at $z=0$, propagates along the fiber, and is gradually coupled into and absorbed by the core. 
The unabsorbed portion of the pump that reaches the output reflector at $z=L$ is reflected back with 100\% efficiency ($R_2(\lambda_p)$=1).  For all our simulations, 
the reflection coefficient at the signal wavelength of the input mirror at $z=0$ is 100\%, i.e.,  $R_1(\lambda_s)=1.0$. The reflection coefficient of the output 
coupler at the signal wavelength, i.e., $R_2(\lambda_s)$, will be optimized for the best performance in our simulations. Note that the material and 
dopant properties of the fiber studied for this simulation are exactly the same as the fiber characterized in the previous sections; however, its geometry is different and customized 
for high-power operation as shown in Fig.~\ref{fig:laser}, i.e., a single-mode core and a double-cladding geometry.

\begin{table}[htbp]
\centering
\caption{\bf Yb-doped ZBLAN fiber simulation parameters}
\scalebox{0.8}{
\begin{tabular}{ccc}
 \hline
 Symbol & Parameter & Value \\
 \hline
$d_{co}$ & core diameter & 6.5 \textmu m \\ 
$d_{cl}$ & cladding diameter & 125 \textmu m\\
$\Gamma_s$ & signal power filling factor & 0.89\\
$\Gamma_p$ & pump power filling factor & $2.704 \times 10^{-3}$\\
$N$ & ${\rm Yb}^{+3}$ concentration & $1.453 \times 10^{26} {\rm m}^{-3}$\\
$\tau$ & upper manifold lifetime & 1.7 ms\\
H & convective heat transfer coefficient & $30\,{\rm W/m}^2 {\rm K}$\\
$\sigma_{\rm abs}(975\,{\rm nm})$ & absorption cross section & $1.28 \times 10^{-24} {\rm m}^{-2}$\\
$\sigma_{\rm em}(975\,{\rm nm})$ & emission cross section & $1.28 \times 10^{-24} {\rm m}^{-2}$\\
$\lambda_f$ & mean florescence wavelength & 994.96 nm\\
$\alpha_b$ & background absorption (pump \& signal) & $4.278 \times 10^{-2} {\rm m}^{-1}$\\
$\eta_q$ & external quantum efficiency & $99.6 \%$ \\
$R_2(\lambda_p)$ & pump reflection of output mirror & $100 \%$\\
$R_1(\lambda_s)$ & signal reflection of input mirror & $100 \%$\\
\end{tabular}}
  \label{tab:values}
\end{table}

The fraction of the pump power coupled to the active core of the fiber is represented by the power filling factor $\Gamma_p$, which is assumed to be the ratio of the 
area of the active core to the area of the multimode inner cladding. The generated and amplified laser signal power, ${P_s}^{\pm}(z)$, is concentrated mainly in the core
with the power filling factor (core overlap factor) $\Gamma_s$. Our analysis is based on the commonly used rate equation model for Yb-doped fibers~\cite{kelson1998strongly}.  
We also consider the background absorption losses to be the same ($\alpha_b$) for both the signal and the pump. 
For continuous wave (CW) lasers, the set of coupled time-independent steady-state rate equations and pump/signal propagation equations are given by:
\begin{align}
\frac{N_2(z)}{N}&=
\dfrac{\Gamma_s \sigma^a_{s} \lambda_s \widetilde{{P}_s\,} (z) + \Gamma_p \sigma^a_{p} \lambda_p  \widetilde{{P}_p} (z)}
{\Gamma_s \sigma^{ae}_{s} \lambda_s \widetilde{{P}_s\,} (z) + \Gamma_p \sigma^{ae}_{p} \lambda_p \widetilde{{P}_p} (z) + h c A\tau^{-1}},\\
\pm \frac{d {P_p}^{\pm}}{dz} &= - \Gamma_p [\sigma^a_{p} N - \sigma^{ae}_{p} N_2(z)] {P_p}^{\pm}(z) - \alpha_b {P_p}^{\pm}(z),\\
\pm \frac{d {P_s}^{\pm}}{dz} &= -  \Gamma_s [\sigma^a_{s} N - \sigma^{ae}_{s} N_2(z)] {P_s}^{\pm}(z) - \alpha_b {P_s}^{\pm}(z),
\end{align}
where we have used the following definitions:
\begin{align}
\widetilde{{P}_s\,} (z) &:= {P_s}^+ (z) + {P_s}^- (z),\qquad \sigma^{ae}_{s} := \sigma^a_{s} + \sigma^e_{s},\\
\nonumber
\widetilde{{P}_p} (z) &:= {P_p}^+ (z) + {P_p}^- (z),\qquad \sigma^{ae}_{p} := \sigma^a_{p} + \sigma^e_{p}.
\end{align}
Here, $N_2(z)$ is the upper manifold population, which varies along the fiber, $N$ is the total ${\rm Yb}^{+3}$ concentration, 
$\lambda_s (\lambda_p)$ is the signal (pump) wavelength, $\sigma^a_{s} (\sigma^a_{p})$ is the absorption cross section at the 
signal (pump) wavelength, $\sigma^e_{s} (\sigma^e_{p})$ is the emission cross section at signal (pump) wavelength, $\tau$ is  
the upper manifold lifetime, and $A$ is the cross-sectional area of the core.
The $\pm$ superscripts and coefficients in the pump and signal power propagation equations signify the positive and negative propagation directions, respectively.
We have used the values of the parameters reported in Table~\ref{tab:values} in our simulations. The value of  ${\rm Yb}^{+3}$ concentration in Table~\ref{tab:values}
is consistent with our measured value of $\alpha_r$. 

To achieve the anti-Stokes fluorescence cooling (and consequently the RBL condition), the laser must be pumped at a wavelength longer than the mean fluorescence 
wavelength ($ \lambda_p > \lambda_f$). We measured the $\lambda_f$ in our Yb-doped ZBLAN fiber sample to be 994.96\,nm. In conventional fiber lasers, the fiber 
is pumped at the wavelength corresponding to the peak of the absorption, which is approximately 975\,nm. For the RBL design, the pump wavelength is considerably 
longer at which the pump absorption coefficient is significantly reduced. Therefore, the optimum design parameters for an RBL system, specifically $L$ and $R_2(\lambda_s)$,
would have to be quite different from that of a conventional fiber laser. An important metric for the performance of a high-power laser is its efficiency;
therefore, we compare the maximum efficiency achievable by the laser system pumped at $\lambda_p$ with the signal operating at $\lambda_s$.

\begin{figure}[h]
\begin{centering}
\includegraphics[width=\columnwidth]{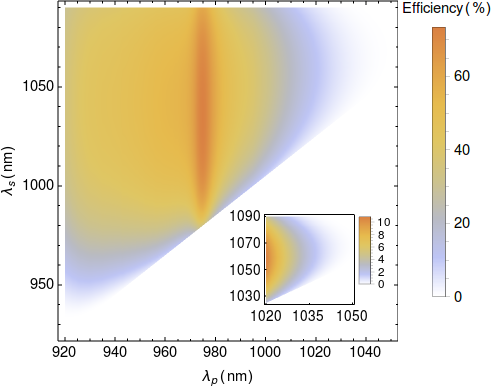}
\caption{The density plot of the {\em optimum} efficiency of the fiber laser for different pump and signal wavelengths, when the laser is pumped with 80\,W of input pump power.
The inset is a magnification of the density plot over the range of wavelengths, which are most relevant for an RBL system.}
\label{fig:efficiency}
\end{centering}
\end{figure}

Figure~\ref{fig:efficiency} shows a density plot of the {\em optimum} efficiency as a function of the pump and signal wavelengths. The efficiency is
defined as the output signal power divided by the input pump power. At every point in Fig.~\ref{fig:efficiency}
identified by a $(\lambda_p,\lambda_s)$ pair, we run an optimization code to find the length $L$ and the output signal reflectivity $R_2(\lambda_s)$ corresponding to the
maximum achievable output signal power for 80\,W of input pump power. Therefore, the efficiency at each point is the maximum value that is achievable in a system design.
Of course, the values of $L$ and $R_2(\lambda_s)$ can be widely different at different points in Fig.~\ref{fig:efficiency}. 
For example, the optimum length and reflectivity for $\lambda_p=1035\,{\rm nm}$ and $\lambda_s=1080\,{\rm nm}$ are 24.1\,m and 30\%, but the same parameters for 
$\lambda_p=940\,{\rm nm}$ and $\lambda_s=1060\,{\rm nm}$ are 8.2\,m and 4\%. Figure~\ref{fig:efficiency} provides a powerful comparison of the maximum achievable 
efficiencies at different wavelengths, where one need not worry about whether the lower efficiency is because of the choice of the wavelengths or because of
the non-optimal choice of the fiber length and output reflectivity. This method of comparison with full optimization is absolutely necessary, because optimum 
RBL design parameters can be very different from those of conventional fiber lasers.

According to the results presented in Fig.~\ref{fig:efficiency}, the maximum efficiency of around 70\% is obtained for the pump wavelength of around 975\,nm, which corresponds to the
peak of the absorption cross section of the Yb-doped ZBLAN. In RBL systems, the pump wavelength is typically longer than 1020\,nm. The inset in Fig.~\ref{fig:efficiency} is 
a magnification of the density plot over the range of wavelengths, which are most relevant for an RBL design. It can be seen that a range of 5-10\% efficiency is the best
that can be achieved in the Yb-doped ZBLAN fiber explored here. These general observations persist for any Yb-doped optical fiber in an RBL design: the maximum achievable
efficiency is reduced as a trade-off for the better heat-management or the total heat mitigation in a strictly RBL system.    

\begin{figure}[h]
\begin{centering}
\includegraphics[width=3.4 in]{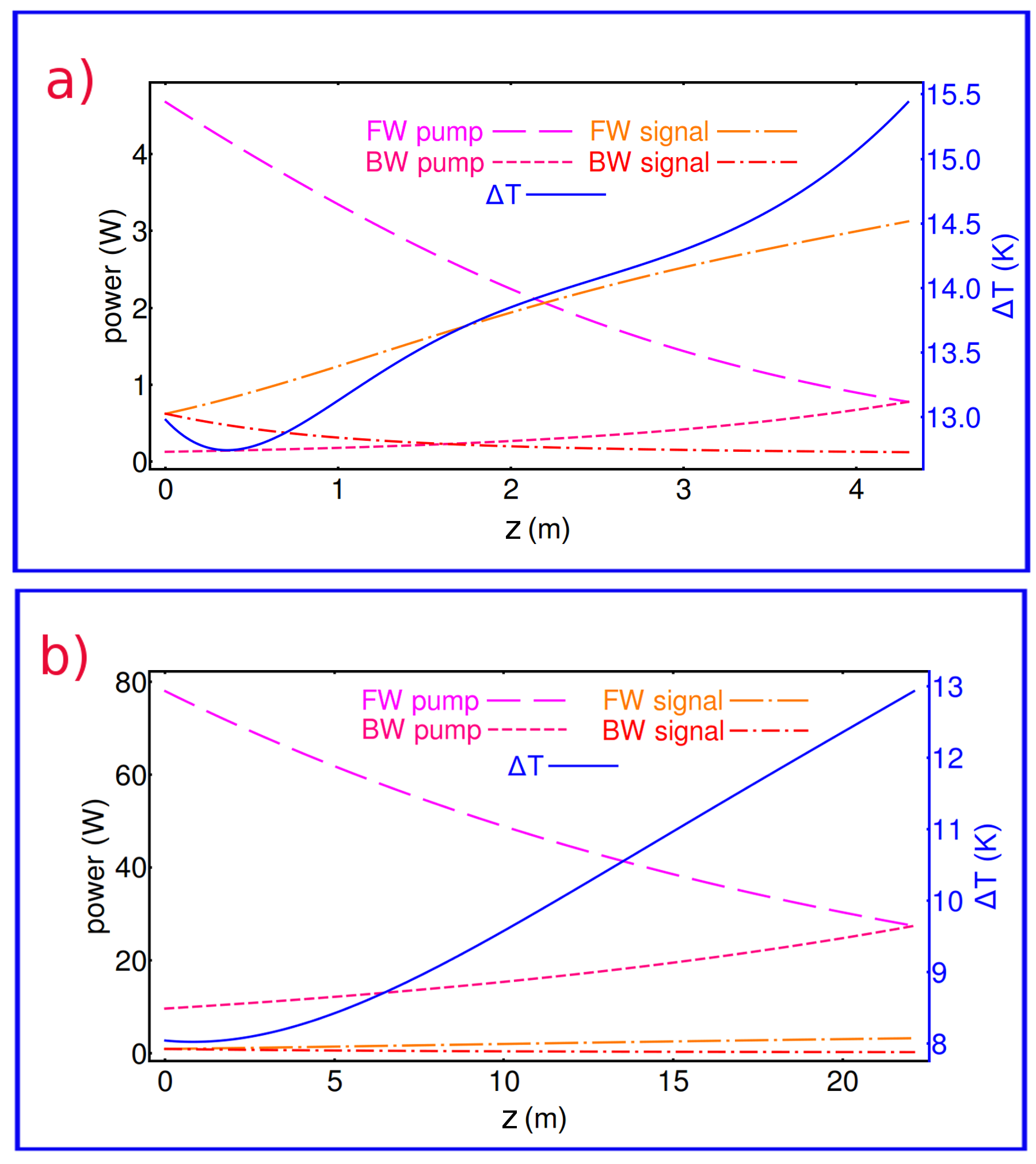}
\caption{a) Propagation of the forward pump (FW pump), backward pump (BW pump), forward signal (FW signal) backward signal (BW signal), and 
temperature rise along the ZBLAN fiber for a conventional fiber laser pumped at $\lambda_p= 975\,{\rm nm}$. b) Similar graph for the RBL operation
pumped at pumped at $\lambda_p= 1030\,{\rm nm}$. Both lasers are optimized for the signal output power of 3\,W at $\lambda_s= 1070\,{\rm nm}$
for $\alpha_b$ from Table~\ref{tab:values}. Note that the fiber in the RBL design is considerably longer than the conventional design.}
\label{fig:temp}
\end{centering}
\end{figure}

Now that we have an account of the maximum achievable efficiency from Fig.~\ref{fig:efficiency}, we still need to answer the key question for an RBL system design:
is it possible to have a fiber RBL with the same output power as a conventional fiber laser, and with the additional benefit of generating little or no heat?
This is the central question in designing a meaningful fiber RBL. To explore this, we consider two simulations: first for a conventional fiber laser pumped
at $\lambda_p= 975\,{\rm nm}$ presented in Fig.~\ref{fig:temp}a, and second an RBL fiber laser pumped at $\lambda_p= 1030\,{\rm nm}$ presented in Fig.~\ref{fig:temp}b.
The output signal in both lasers is generated to be at $\lambda_s= 1070\,{\rm nm}$ and is 3\,W in power. Both systems are optimized for maximum efficiency. 
Of course, for the conventional fiber laser corresponding to Fig.~\ref{fig:temp}a, the required pump power is only 4.68\,W because of the higher efficiency.
However, for the RBL design, we need 78\,W of input pump power to achieve 3\,W of output signal power due to the lower efficiency.
In each plot, we show the temperature rise ($\Delta T$) as a function of $z$ along the fiber. 
It can be seen that the temperature rise in both designs is comparable, i.e., although the RBL design is pumped at a wavelength at which ASF cooling is 
mitigating some of the generated heat, its lower efficiency is resulting in the same level of temperature rise. This means that we cannot obtain any meaningful 
design because the best we can achieve is the same level of temperature rise albeit with a substantially higher pump power. A slightly better heating performance 
in the RBL design does not justify a 20-fold increase in the pumping power.   

\begin{figure}[h]
\begin{centering}
\includegraphics[width=3.4 in]{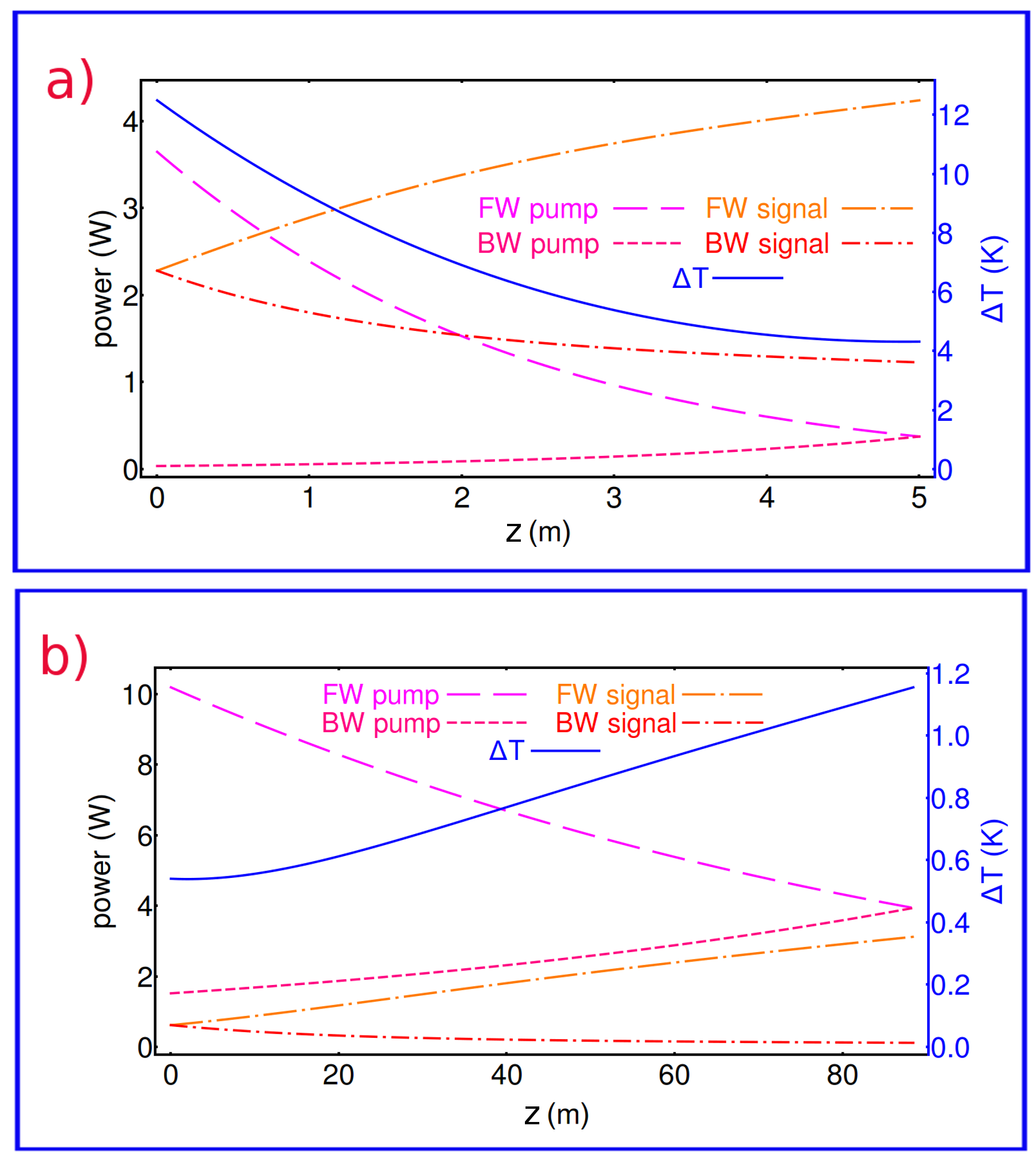}
\caption{Similar to Fig.~\ref{fig:temp}, except the ZBLAN fiber is chosen with a 10-fold reduction in the background absorption, i.e. $\alpha_b^\prime=\alpha_b/10$.
This reduced value is used for both the conventional laser in subfigure a, pumped with 3.65\,W at $\lambda_p= 975\,{\rm nm}$; and the RBL laser in subfigure b,
pumped with 10.2\,W at $\lambda_p= 1030\,{\rm nm}$. 
Both lasers are optimized for the signal output power of 3\,W at $\lambda_s= 1070\,{\rm nm}$. 
The RBL design has a substantially reduced temperature
performance compared with the conventional laser the trade-off of a nearly 2-fold increase in the required pump power.}
\label{fig:temp2}
\end{centering}
\end{figure}

We have identified the large value of the background absorption $\alpha_b$ as shown in Table~\ref{tab:values} to be the root cause of 
making such RBL designs pointless as shown in Fig.~\ref{fig:temp}. Therefore, the key to achieving a viable RBL system is to lower the value of the parasitic 
background absorption, which should be tackled by the proper fabrication and composition of the glass. To support this argument,
we have repeated the simulation presented in Fig.~\ref{fig:temp} with a 10-fold reduction in the background absorption, i.e. $\alpha_b^\prime=\alpha_b/10$~\cite{knall2019demonstration}.
In fact, this a totally reasonable assumption, considering the fact that the value of $\alpha_b$ that we measured earlier for the ZBLAN fiber appears to be too
high, most likely due to the age of the sample and exposure to moisture and oxygen. 
After optimizing the cavity and finding the best reflector and length of the fiber for 3\,W of output signal power, we show the designs in Fig.~\ref{fig:temp2}.
Fig.~\ref{fig:temp2}a which corresponds to the conventional laser is slightly cooler than the design in Fig.~\ref{fig:temp}a; however, the RBL design in  
Fig.~\ref{fig:temp2}b shows much smaller temperature rise compared with Fig.~\ref{fig:temp}b, hence confirming our claim.
\section{Discussion and Conclusion}
Anti-Stokes fluorescence cooling and radiation-balancing of fiber lasers and amplifiers rely on the availability of cooling-grade rare-earth-doped optical fibers. 
Herein, we have performed a detailed investigation of a Yb-doped ZBLAN optical fiber to assess its gain, loss, and cooling-related parameters. The techniques 
are based on the LITMoS test developed in Sheik-Bahae's research group and the MACSLA method recently developed in our group and give accurate results 
on the cooling behavior of the Yb-doped ZBLAN fiber. A main advantage of the MACSLA method is that unlike the cutback method, it is not destructive. We emphasize 
that this work constitutes the first such detailed assessment of a cooling-grade optical fiber over a range of wavelengths. More importantly, the entire measurement
is performed in atmospherics pressure using conventional table-top optical equipment. It is shown that the specific ZBLAN fiber sample measured in the experiments 
is cooling-grade. However, its parasitic background absorption is too large to be used for proper radiation-balancing in a fiber laser or amplifier design.

As it is shown in Fig.~\ref{fig:efficiency}, the maximum efficiency for our Yb-doped ZBLAN fiber is obtained when it is pumped around 975\,nm wavelength, while the efficiency 
of the laser in favorable wavelengths for RBL operation ($\lambda_p \gtrsim 1020\,{\rm nm}$) is $\sim$7 times smaller. This lower efficiency is mainly due to 
the smaller absorption cross section of the pump power at RBL pump wavelengths. In order to asses the viability of an RBL design, we compared a conventional
laser pumped at 975\,nm wavelength with an RBL design pumped at $\lambda_p= 1030\,{\rm nm}$ (see Fig.~\ref{fig:temp}). The output signal in both lasers was assumed to be
3\,W in power at $\lambda_s= 1070\,{\rm nm}$ and both systems were optimized for maximum efficiency (fiber length and output signal reflectivity). We observed 
that the temperature rise in both designs were comparable, while the RBL design required a 20-fold larger pump power, making the RBL design totally unpractical.
We attributed the problem to the relatively large value of the parasitic absorption of the pump $\alpha_b$; in a separate simulation in Fig.~\ref{fig:temp2},
we showed that a 10-fold reduction in $\alpha_b$ can reduce the heating in the RBL design significantly and make it viable. Therefore, we conclude that the
key to achieving a viable RBL design in optical fibers is to focus on the fabrication and composition of the glass to reduce $\alpha_b$. For fiber lasers, because 
the parasitic heat is proportional to $\alpha_b$ and the pump/signal powers, and the fact that the heat is all dumped in small core/cladding cross sections, 
the demand on reducing $\alpha_b$ is higher than other forms of RBL lasers. In fact, our analysis has shown that while the value of $\alpha_b$ is small enough 
to make our sample a cooling-grade fiber, it is not sufficiently small to make it a viable gain medium for RBL operation. Therefore, as far as fiber lasers are concerned,
a viable RBL laser is more demanding on the parasitic absorption coefficient than the laser cooling experiment.
\section{Appendix: Derivation of the Cooling Efficiency Formula}
The cooling efficiency, $\eta_c$, is defined as the net power density (per unit volume) extracted from the material ($P_{\rm net}$) 
per unit power density absorbed or scattered ($P_{\rm abs}$): $\eta_c=P_{\rm net}/P_{\rm abs}$. We can write $P_{\rm net}=P_{\rm asf}-P_{\rm abs}$, where
$P_{\rm asf}$ is the fraction of the ASF power density that escapes the cooling material. The absorbed power density is given by $P_{\rm abs}=(\alpha_r+\alpha_b)I_P$,
where $I_P$ is the pump intensity, $\alpha_r$ is the resonant absorption of the pump laser due to the presence of the gain materials (Yb ions in here), 
$\alpha_b$ represents the parasitic background absorption and scattering of the pump laser. The ASF power density is therefore given by $\eta_e N_2 W_r (h\nu_f)$, 
where $\nu_f$ is the mean florescence frequency, $N_2$ is the number density of the excited upper level in the quasi two-level Yb ions, and 
$W_r$ ($W_{nr}$) is the radiative (non-radiative) decay rate of the excited state of the doped ions. $\eta_e$ is the extraction (escape)
efficiency and $1-\eta_e$ is the fraction of photons which are radiated but are trapped inside the host. The rate equation can be expressed as
\begin{align}
\dfrac{dN_2}{dt}=\dfrac{\alpha_r I_P}{h\nu_p}-(W_r+W_{nr})N_2+(1-\eta_e)W_rN_2,
\end{align}
where we have assumed that the trapped florescence is reabsorbed by the Yb ions. In steady-state, where $dN_2/dt=0$, we can solve for $N_2$ and obtain 
$P_{\rm asf}=\alpha_r I_P\eta_q(\lambda_p/\lambda_f)$, where the external quantum efficiency is given by $\eta_q=\eta_eW_r/(\eta_eW_r+W_{nr})$,
and $\lambda_p$ ($\lambda_f$) is the pump (mean florescence) wavelength. We therefore have
\begin{align}
P_{\rm net}=(\alpha_r+\alpha_b)I_P-\alpha_r I_P\eta_q(\lambda_p/\lambda_f).
\end{align}
We can use these results to present the cooling efficiency in the form of Eq.~\ref{eq:etac}.
\section{Methods: Cooling grade polishing of the ZBLAN fiber}
The laser cooling and LITMoS test are very sensitive to surface impurities: dust particles, contamination, or scratches on the fiber facets, which act as sources of external heating 
and can negatively impact the laser cooling experiment. Therefore, high-quality polishing and cleaning of the optical fiber facets are critical steps for 
a successful LITMoS experiment~\cite{seletskiy2010laser,peysokhan2018measuring}. ZBLAN fibers are made from a soft glass, which is prone to oxidation; therefore, it is much harder 
to polish their facets to a high-quality finish compared with silica fibers. Moreover, commercial equipment for polishing and processing ZBLAN fibers are not as widely available as 
for silica fibers. In this section, we detail the procedure we followed to polish our ZBLAN fiber for the laser cooling experiment.

\begin{figure}[h]
\begin{centering}
\includegraphics[width=\columnwidth]{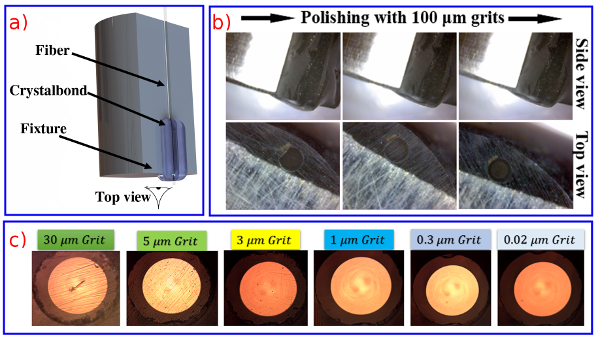}
\caption{a) Images of the polishing fixture and the ZBLAN doped fiber, which are glued together by the \emph{Crystalbond}. 
b) Initial coarse polishing steps to prepare a flat surface for further polishing. From left to right, the side and facet views of 
the doped fiber are shown for each step of the coarse polishing. c) Images of the facet of the ZBLAN fiber under microscope after 
each fine polishing step.}
\label{fig:polishing}
\end{centering}
\end{figure}
To begin the polishing process, as it is shown in Fig.~\ref{fig:polishing}(a), we placed the fiber on a home-made polishing fixture, which is made of stainless steel. 
The choice of a stainless steel polishing fixture is crucial because the aluminum oxide that forms on an aluminum fixture can easily delaminate 
during the polishing procedure and can scratch the fiber facet.
We heated \emph{Crystalbond 509} and used it glue the fiber in its position in the polishing fixture. The fiber and \emph{Crystalbond} were then  
polished with a 30\,\textmu m grit polishing sheet. As it is shown in Fig~\ref{fig:polishing}(b), this procedure continues until the 
full circular shape of the core-cladding of the facet of the fiber appears under the microscope. We followed a 
wet polishing procedure and used water-free \emph{glycol} and \emph{glycerin} combination for the liquid element because 
ZBLAN interacts with \emph{oxygen} and \emph{OH}. The polishing procedure continued with polishing sheets of 5, 3, 1, 0.3 and 0.02\,\textmu m grits, sequentially. 
After each step, the facet of fiber was inspected under a microscope and if there was a scratch, we repeated the previous steps. The final surface of the doped ZBLAN 
fiber is shown in the last image of Fig.~\ref{fig:polishing}(c). 
\begin{figure}[h]
\begin{centering}
\includegraphics[width=\columnwidth]{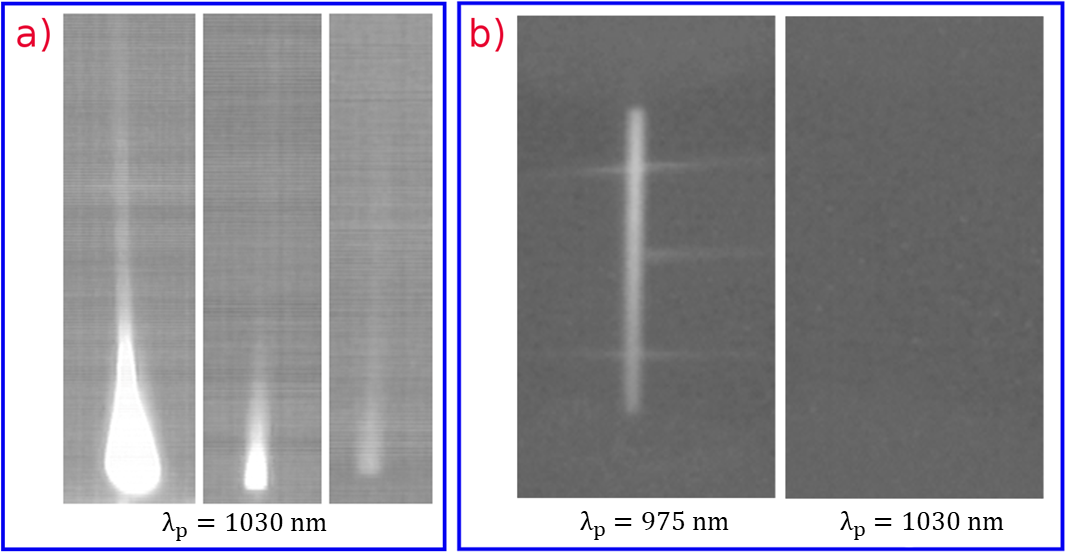}
\caption{Thermal camera images of the laser-pumped ZBLAN fiber. Images in subfigure (a) are for pumping at 1030\,nm wavelength and sequentially 
improved polishing of the facets. The brighter spots indicate heating and as the polishing quality is improved, the facet heating is reduced. 
When cooling-grade polishing is reached the facets no longer are sources of parasitic heating in subfigure (b), transition from heating to cooling
is clearly observed when the pump wavelength is switched from 975\,nm to 1030\,nm wavelength.}
\label{fig:fiber}
\end{centering}
\end{figure}

Cleaning the facets is another crucial step in fiber preparation. The cleaning is done immediately after the fiber is polished. The fiber and the polishing fixture 
are immersed in a 99.5\% \emph{Acetone} solution in an ultrasound bath for about 30 minutes for dissolving the bulk \emph{Crystalbond} and detaching the fiber from the polishing fixture. 
The fiber is then soaked in a \emph{Citrosolve} solution in the ultrasound bath to dissolve the residual \emph{Crystalbond} for about one hour. The fiber is then dipped in 
a 99.5\% \emph{Acetone} solution for about 4 hours to clean the fiber thoroughly. Following this, the fiber is immersed in the 99.999\% \emph{Isopropanol Alcohol} for 30 minutes 
in the ultrasound bath to clean any trace of \emph{Aceton}. To emphasize the importance of such a cooling grade polishing procedure, in Fig.~\ref{fig:fiber}(a) we show the
thermal camera images of the ZBLAN fiber that is pumped at 1030\,nm wavelength. The 3 images show, sequentially, that the improved polishing of the facets results in reduced
heating of the facets. In Fig.~\ref{fig:fiber}(b) where the fiber is highly polished and the facets no longer are sources of parasitic heating, one can clearly observe 
the transition from heating when the fiber is pumped at 975\,nm wavelength to cooling when pumped at 1030\,nm wavelength.
\section*{Acknowledgments}
The authors would like to acknowledge
M. Sheik-Bahae, R. I. Epstein, M. P. Hehlen, M. Hossein-Zadeh and A. R. Albrecht for informative discussions. 
This material is based upon work supported by the Air Force Office of Scientific Research under
award number FA9550-16-1-0362 titled Multidisciplinary
Approaches to Radiation Balanced Lasers (MARBLE).

\end{document}